# Challenges In Extracting Charged Current Quasi-Elastic Model Information in MiniBooNE's Anti-Neutrino Data

Joseph Grange[a] for the MiniBooNE collaboration

[a]*University of Florida, Gainesville FL*

**Abstract.** Using a high-statistics sample of anti-neutrino charged current quasi-elastic (CCQE) events, MiniBooNE reports the challenges in measuring parameters within the Relativistic Fermi Gas model. As the CCQE analysis has been completed in MiniBooNE's neutrino data, particular attention is paid to the differences in CCQE interactions between the two running modes.

## CCQE MODEL TUNING AT MINIBOONE

At MiniBooNE's sub-GeV energies, interaction rates are dominated by the CCQE cross section. In neutrino mode, describing these interactions well was an essential constraint in the oscillation search to either refute or confirm the controversial LSND results[1]. Two parameters in the Relativistic Fermi Gas (RFG) model were simultaneously tuned to improve data - Monte Carlo agreement[2]. The first is the axial mass form factor $M_A$ ( $\sigma = \sigma(F_A)$ ):

$$F_A(Q^2) = g_A \bigg/ \left(1 + \frac{Q^2}{M_A^2}\right)^2 \tag{1}$$

where $Q^2$ is the four momentum transfer and $g_A = 1.267$ is well known from beta decay experiments. MiniBooNE assumes a dipole form for $F_A(Q^2)$. Tuning $M_A$ alone left a data deficit in the low $Q^2$ region, so a pauli-blocking parameter $\kappa$ was introduced and fit together with $M_A$. The scale factor $\kappa$ controls the lower bound on the available nucleon energy:

$$E_{lo} = \kappa\left(\sqrt{p_F^2 + M^2} - \omega + E_B\right) \tag{2}$$

where $E_B$ is the binding energy of the nucleon, M is the mass of the nucleon, $p_F$ the Fermi momentum, and $\omega$ the energy transfer.

MiniBooNE measured $M_A = 1.35 \pm 0.17$ GeV and $\kappa = 1.007\,^{+\,0.007}_{\,-\,\infty}$ for sub-GeV neutrino CCQE events on a CH_2 target[3].

## INTERACTIONS FROM MINIBOONE'S ANTI-NEUTRINO BEAM

The neutrino mode CCQE analysis benefitted from both a low wrong-sign anti-neutrino background and a single nuclear scattering target. Unfortunately the same is not true in the anti-neutrino beam. The wrong sign neutrino contamination is more significant and the anti-neutrinos scatter off bound nucleons in carbon and free nucleons in hydrogen. The interaction is described in Figure 1, while Table 1 summarizes the prediction for MiniBooNE contamination and free nucleon scattering based on MC simulations. We use the NUANCE neutrino interaction generator[4], which uses the RFG model[5] for CCQE scattering[6].

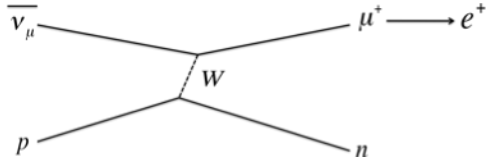

**FIGURE 1.** An anti-neutrino CCQE interaction. MiniBooNE reconstructs the event based on kinematic measurements of the outgoing muon.

**TABLE 1.** Summary of wrong sign contamination and free nucleon scattering in MiniBooNE's two running modes, based on Monte Carlo simulation.

| Component | Neutrino Beam | Anti-Neutrino Beam |
| --- | --- | --- |
| Right Sign CCQE | 77 % | 54 % |
| Wrong Sign CCQE | 2 % | 22 % |
| QE H2 Scattering | 0 % | 19 % |

The wrong sign contamination will be measured from the data and subtracted from the CCQE signal, but the QE model parameter measurement is not so straightforward. While the parameter $\kappa$ measured from the neutrino beam interactions is consistent with 1 and consequently perhaps not strictly needed, it is less clear what is the correct approach to take with regard to the axial mass $M_A$. In the past $M_A$ was thought to be well known at an appreciably lower value than what MiniBooNE measured. Based on bubble chamber data, $M_A$ was measured to be $1.03 \pm 0.02$ GeV[7]. Other neutrino experiments besides MiniBooNE have also measured higher values of $M_A$ (see Table 2).

**TABLE 2.** Some $M_A$ measurements from recent neutrino experiments, most significantly higher than the previous bubble chamber measurement of $1.03 \pm 0.02$ GeV[7].

| Source | Measured $M_A$ (GeV) |
| --- | --- |
| K2K SciFi[8] | $1.20 \pm 0.12$ |
| K2K SciBar[9] | $1.14 \pm 0.11$ |
| MINOS[10] | $1.26 \pm 0.17$ |
| NOMAD[11] | $1.07 \pm 0.07$ |
| MiniBooNE[3] | $1.35 \pm 0.17$ |

Within the RFG model, it is possible that $M_A$ is a nuclear dependent parameter and may also be a function of the incident neutrino energy. We should accommodate this possibility and not strictly assume a single $M_A$ for all of our anti-neutrino interactions. Of course it is also possible nuclear effects such as charge exchange/absorption and final state interactions are obfuscating the various measurements of $M_A$. Regardless, when fixing $M_A$ and $\kappa$ at their previous "world values" of 1.02 GeV and 1.000 respectively, MiniBooNE sees poor data - Monte Carlo agreement in the kinematic variables. See Figures 2 and 3.

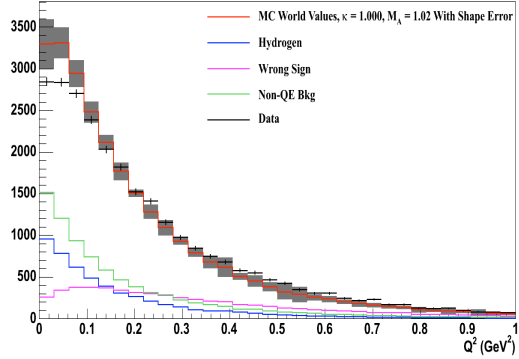

**FIGURE 2.** Reconstructed four-momentum transfer $Q^2$ with QE model parameters evaluated at the world values $M_A = 1.02$ GeV, $\kappa = 1.000$ compared to data. Significant shape discrepancy at both low and high $Q^2$ is observed. The complicating backgrounds are plotted. Distributions have arbitrary units.

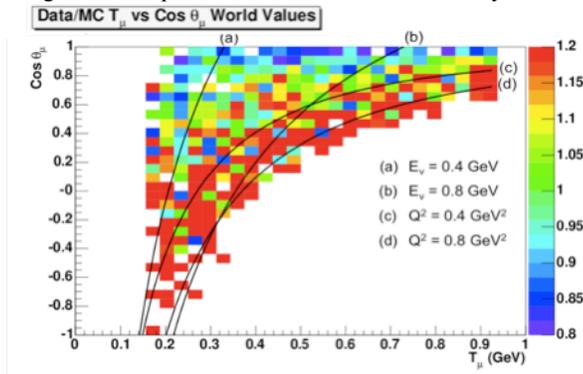

**FIGURE 3.** Ratio of data / Monte Carlo in kinetic energy $T_\mu$ vs. cosine of the scattering angle Cos $\theta_\mu$ of the outgoing muon for $M_A = 1.02$ GeV, $\kappa = 1.000$. Discrepancies follow lines of constant $Q^2$ and not $E_\nu$. This suggests the disagreement may originate in the cross section and not the flux prediction.

However, if we apply $M_A = 1.35$ GeV and $\kappa = 1.007$, the values measured in MiniBooNE's neutrino mode, we see improved data - Monte Carlo agreement. See Figures 4 and 5. The axial mass $M_A = 1.35$ GeV is applied to all QE events, regardless of sign and scattering target.

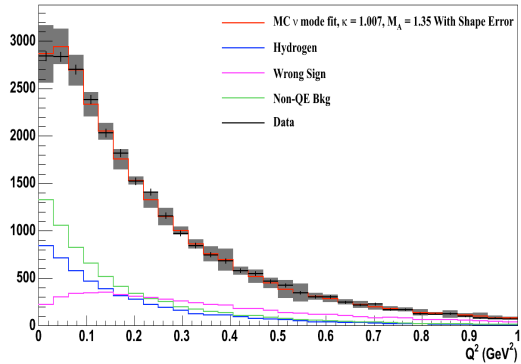

**FIGURE 4.** Reconstructed four-momentum transfer $Q^2$ with QE model parameters evaluated at the neutrino mode best fit values compared to data. The shape disagreement observed when using the world values of $M_A$ and $\kappa$ is improved. The complicating backgrounds are plotted. Distributions have arbitrary units.

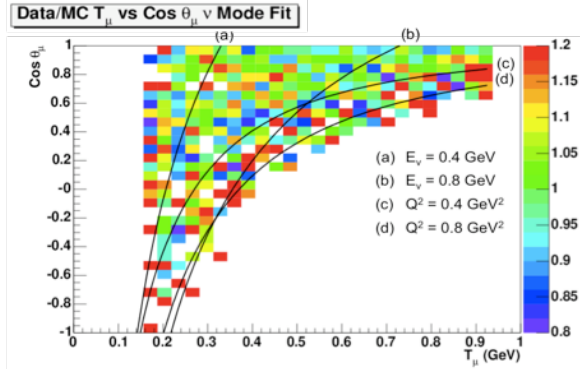

**FIGURE 5.** Ratio of data / Monte Carlo in kinetic energy $T_\mu$ vs. cosine of the scattering angle $\cos\theta_\mu$ of the outgoing muon for $M_A = 1.02$ GeV, $\kappa = 1.000$. The systematic disagreement observed when using $M_A$ and $\kappa$ world values is improved by applying those measured in MiniBooNE's neutrino mode.

We do not mean to suggest the resolution to the task of measuring CCQE values in anti-neutrino mode is to blindly apply those found in neutrino mode. Rather, the improved data - Monte Carlo agreement when using the neutrino mode values suggests that pursuing to "fix" the discrepancy through the measurement of $M_A$ and $\kappa$ is promising.

## ACKNOWLEDGMENTS

We thank the organizers of the NuInt 2009 conference.